\documentclass{aa}  

\usepackage{graphicx}
\usepackage{txfonts}
\usepackage{lscape}
\usepackage{tabularx}
\usepackage{adjustbox}
\usepackage{comment}
\usepackage{hyperref}
\begin{document} 
\newcommand{\deufrac}{\smash{$D_{\rm frac}^{\rm N_2H^+}$}}

   \title{Deuterium Fractionation across the Infrared Dark Cloud G034.77$-$00.55 interacting with the Supernova Remnant W44}

   %\subtitle{}

   \author{G. Cosentino\inst{1}\thanks{E-mail:giuliana.cosentino@chalmers.se},
    J. C. Tan\inst{1,2}\fnmsep,
    I. Jim\'{e}nez-Serra\inst{3},
    F. Fontani\inst{4},
    P. Caselli\inst{5},
    J. D. Henshaw\inst{6,7},
    A. T. Barnes\inst{8},
    C.-Y. Law\inst{1,8},
    S. Viti\inst{9,10},
    R. Fedriani\inst{1,11},
    C.-J. Hsu\inst{1},
    P. Gorai\inst{1},
    S. Zeng\inst{12}}
    \authorrunning{Cosentino et al.}
    \titlerunning{Deuterium Fractionation in the IRDC G34.77 interacting with the SNR W44.}
   \institute{Department of Space, Earth and Environment, Chalmers University of Technology, SE-412 96 Gothenburg, Sweden
    \and 
    Department of Astronomy, University of Virginia, 530 McCormick Road Charlottesville, 22904-4325 USA
    \and
    Centro de Astrobiolog\'{i}a (CSIC/INTA), Ctra. de Torrej\'on a Ajalvir km 4, Madrid, Spain
    \and
    INAF  Osservatorio Astronomico di Arcetri, Largo E. Fermi 5, 50125 Florence, Italy
    \and
    Max Planck Institute for Extraterrestrial Physics, Giessenbachstrasse 1, 85748 Garching bei M\"{u}nchen, Germany
    \and
    Astrophysics Research Institute, Liverpool John Moores University, 146 Brownlow Hill, Liverpool L3 5RF, UK
    \and
    Max-Planck-Institut f\"ur Astronomie, K\"onigstuhl 17, D-69117 Heidelberg, Germany
    \and
    European Southern Observatory, Karl-Schwarzschild-Strasse 2, D-85748 Garching, Germany
    \and
    Leiden Observatory, Leiden University, PO Box 9513, 2300 RA Leiden, The Netherlands
    \and
    Department of Physics and Astronomy, University College London, Gower Street, London, WC1E 6BT, UK
    \and
    Instituto de Astrof\'isica de Andaluc\'ia, CSIC, Glorieta de la Astronom\'ia s/n, E-18008 Granada, Spain
    \and
    Star and Planet Formation Laboratory, Cluster for Pioneering Research, RIKEN, 2-1 Hirosawa, Wako, Saitama, 351-0198, Japan}
   \date{Received ---; accepted ---}

% \abstract{}{}{}{}{} 
% 5 {} token are mandatory
 
  \abstract
   {Supernova remnants (SNRs) may regulate star formation in galaxies. For example, SNR-driven shocks may form new molecular gas or compress pre-existing clouds and trigger the formation of new stars.}
   {To test this scenario, we measure the deuteration of $\rm N_2H^+$, \deufrac, a well-studied tracer of pre-stellar cores, 
   across the Infrared Dark Cloud (IRDC) G034.77-00.55, known to be experiencing a shock interaction with the SNR W44.}
   {We use N$_2$H$^+$ and N$_2$D$^+$ J=1-0 single pointing observations obtained with the 30m antenna at the Instituto de Radioastronomia Millimetrica to infer \deufrac toward five positions across the cloud, namely a massive core, different regions across the shock front, a dense clump and ambient gas.}
   {We find \deufrac in the range 0.03-0.1, several orders of magnitude larger than the cosmic D/H ratio ($\sim$10$^{-5}$). Across the shock front, \deufrac is enhanced by more than a factor of 2 (\deufrac$\sim$0.05-0.07) with respect to the ambient gas ($\leq$0.03) and similar to that measured generally in pre-stellar cores. Indeed, in the massive core and dense clump regions of this IRDC we measure \deufrac$\sim$0.1.}
   {We find enhanced deuteration of $\rm N_2H^+$ across the region of the shock, at a level that is enhanced with respect to regions of unperturbed gas. It is possible that this has been induced by shock compression, which would then be indirect evidence that the shock is triggering conditions for future star formation. However, since unperturbed dense regions also show elevated levels of deuteration, further, higher-resolution studies are needed to better understand the structure and kinematics of the deuterated material in the shock region, e.g., if it still in relatively diffuse form or already organised in a population of low-mass pre-stellar cores.}


   \maketitle
    
%
%-------------------------------------------------------------------
\section{Introduction}
Infrared Dark Clouds (IRDCs) are dense \cite[$n_{\rm H}\geq10^4$ cm$^{-3}$;][]{butlerTan2012} and cold \cite[T$\leq$20 K;][]{pillai2006} regions of the Interstellar Medium (ISM) that have high column densities and extinctions and thus can appear dark against the diffuse Galactic mid-infrared background. IRDCs are the densest regions of giant molecular clouds (GMCs) and are known to host the formation of stars from low, to intermediate to high-mass \citep[e.g.,][]{tan2013,foster2014, pillai2019,moser2020,yu2020}. Despite their importance for the stellar content of galaxies \citep[e.g.,][]{tan2014,hernandez2015,peretto2016,RetesRomero2020,morii2021}, it is still unclear how star formation is initiated in IRDCs. 

IRDC formation from GMCs mediated by compression due to internal cloud turbulence is one possible scenario \citep[e.g.,][]{krumholz2005}. Formation due to decay of turbulent and/or magnetic support support, perhaps associated with the global collapse of a recently formed GMC, is another proposed model \citep[e.g.,][]{vazquez2011}. Yet another scenario is triggered compression by galactic shear driven GMC-GMC collisions \citep{tan2000}, with various numerical simulation studies of this process carried out by \citep[e.g.,][]{tasker2009,wu2015,wu2017}. The involvement of an external agent to a given GMC, i.e., collision with another GMC, can help explain the large dispersion in star formation activity in GMC populations \citep[e.g.,][]{tan2000,lee2016}. Finally, it has also been proposed that IRDC and star formation are triggered by stellar feedback driven shocks, e.g., HII regions or supernova remnants \citep[e.g.,][]{inutsuka2015}. In systems like the Milky Way, where the global star formation rate (SFR) is relatively low compared to the amount of dense gas available \citep[][]{zuckerman1974,krumholz2007}, stellar feedback driven star formation is unlikely to account for the majority of star formation, since if it did, the SFR would grown exponentially to large values. Nevertheless, the process may occur occasionally and such instances are of great interest, e.g., to see if the star formation process varies when triggered in this way.

There are a number of recent observational studies presenting evidence for instances of IRDC and star formation triggered by SNR-driven shocks \citep[e.g.,][]{cosentino2019,ricovillas2020,cosentino2022}. 
\cite{cosentino2019} investigated how the shock driven by the SNR W44 affects the physical conditions of the interacting cloud, the IRDC G034.77-00.55 \citep[hereafter G34.77 or IRDC G from the sample of][]{butlerTan2012}. Toward this source, the shock interaction is occurring at the edge of the IRDC, i.e., toward an arch-like ridge with no signatures of deeply embedded protostars \citep{cosentino2018, cosentino2019, barnes2021}. By using ALMA images of Silicon monoxide (SiO), \cite{cosentino2019} studied the shocked gas kinematics and inferred a shock velocity of $\sim$20 km s$^{-1}$, compatible with that previously estimated by \cite{sashida2013}, and a shock dynamical age of $\sim1.5\times10^4$ years. From C$^{18}$O observations, the authors estimated a density enhancement caused by the shock of a factor $>10$, i.e., with post-shocked volume densities n(H$_2)>$10$^5$ cm$^{-3}$.

In this paper, we investigate how the SNR-driven shock may be affecting the chemical properties of the molecular gas. A well-studied tracer of the chemical conditions at the onset of star formation is the fraction of Deuterium with respect to Hydrogen (D/H) measured from N$_2$H$^+$ and N$_2$D$^+$ emission (hereafter \deufrac). \deufrac, defined as the ratio of the column densities of the deuterated and non-deuterated species, is highly enhanced in pre-stellar cores \citep{crapsi2005, caselli2008, emprechtinger2009}, i.e., with values $\sim 0.1$, compared to the cosmic D/H abundance \citep[$\sim$10$^{-5}$;][]{oliveira2003}. This is due to a combination of low temperatures ($T\leq$20~K) and high-densities (n(H$_2)\geq10^4$ cm$^{-3}$), typical of the pre-stellar phase, that lead to high levels of CO freeze-out onto dust grains \citep[e.g., ][]{caselli1999}. With the CO largely depleted, reactions between N$_2$ and H$_3^+$ (and its deuterated form H$_2$D$^+$) can efficiently occur, boosting the formation of N$_2$D$^+$ \citep[e.g.,][]{dalgarno1984,caselli2002,walmsley2004}. \deufrac is therefore considered an optimal tracer of the evolutionary phase of both low- and high-mass star forming objects \citep{fontani2011,tan2013,giannetti2019}.

Here we estimate \deufrac in several regions of the IRDC G34.77. To the best of our knowledge, no previous studies have investigated \deufrac in shocks driven by SNRs. However, previous studies have measured D/H in other species in shocks associated with molecular outflows \citep[e.g.,][]{codella2013,fontani2014,busquet2017}. In these environments, the shock passage causes the release into the gas phase of deuterated species whose formation occurs on the icy mantle of dust grains, e.g., deuterated counterparts of CH$_3$OH and H$_2$CO \citep{fontani2014,busquet2017}. The sputtering and grain-grain collision processes enabled in shocks increase the CO abundance in the gas phase and the formation of N$_2$D$^+$ is therefore expected to be suppressed. However, at the high densities reached in the post-shocked gas, gas-phase chemistry can proceed at a much faster rate and the CO depletion timescales can be shortened \citep{lis2002,lis2016}. This may therefore result in enhanced abundances of N$_2$D$^+$.

The Letter is organised as follows. In \S\ref{obs}, we describe the observations and discuss the selection of positions across the cloud. In \S\ref{method}, we describe the method applied to estimate \deufrac. In \S\ref{resDisc}, we report and discuss our results. Finally, in \S\ref{conclusions}, we present our conclusions. 

%--------------------------------------------------------------------
\section{Observations and target selection.}\label{obs}

\begin{figure}[!htpb]
    \centering
    \includegraphics[width=0.5\textwidth,trim=0cm 0cm 0cm 0cm, clip=True]{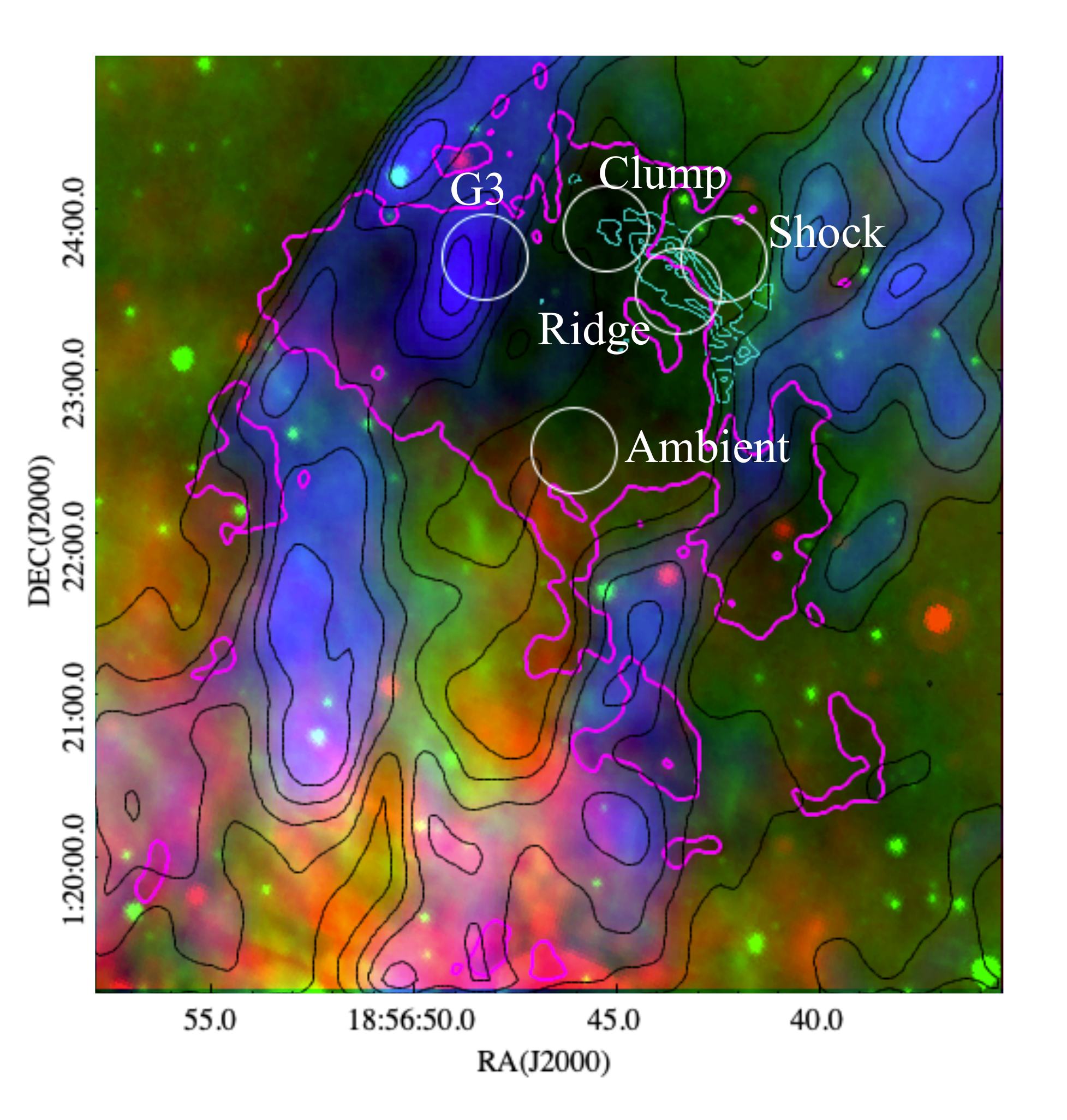}
    \caption{Three colour image of G34.77. Red is 24 $\mu$m emission \citep[Spitzer MIPSGAL;][]{carey2009}, green is 8 $\mu$m emission \citep[Spitzer GLIMPSE;][]{churchwell2009} and blue is 1 GHz continuum emission \citep[THOR survey;][]{beuther2016} (black contours from 3 to 27 Jy beam$^{-1}$ by 6 Jy beam$^{-1}$). The continuum emission probes the gas  associated with the SNR expanding shell. The magenta contour \citep[$A_V=20\:$mag][]{kainulainen2013} highlights the cloud shape. Superimposed on the map are the five positions of interest (white circles; $34^{\prime\prime}$ IRAM30m beam at the N$_2$D$^+$ frequency of 77.112 GHz). The cyan contours correspond to the ALMA SiO contours from \cite{cosentino2019} (from 0.05 Jy beam$^{-1}$ to 0.3 Jy beam$^{-1}$ by 0.05 Jy beam$^{-1}$).}
    \label{fig:fig1}
\end{figure}

In March 2020, we observed the $J=1\rightarrow0$ rotational transitions of N$_2$H$^+$ and N$_2$D$^+$ toward five positions across the IRDC G34.77 (Table~\ref{tab:tabpos}). These positions (white circles) are shown in Figure~\ref{fig:fig1} and listed in Table~\ref{tab:tabpos}. They were selected to probe different environments across G34.77, i.e., the known massive core G3 from \cite{butlerTan2012}, unperturbed gas within the cloud (``Ambient''), a dense clump adjacent to the shock front \citep[``Clump'';][]{barnes2021}, and two regions across the shock front seen by ALMA, i.e., ``Shock'' and ``Ridge'' \citep{cosentino2019}. The core G3, first detected as a point-like source, MM4, in the 1.2 mm images presented by \cite{rathborne2006}, shows no evidence of a central protostellar source. This is indicated by the lack of point-like $8\:\mu$m \citep{rathborne2006}, $24\:\mu$m and $4.5\:\mu$m \citep{chambers2009} sources and SiO emission \citep{cosentino2018}. Toward the Clump, \cite{barnes2021} report no evidence of 3~mm continuum and IR emission, indicating that the region is not harbouring deeply embedded protostars. Therefore, G3 and Clump may be representative of starless regions. However, we note that since 1.2~mm continuum emission is detected toward G3, this may be at an evolutionary stage that is more advanced than the Clump. The Shock position corresponds to the region where the higher velocity shocked material (SiO at $\sim$44-45 km\,s$^{-1}$) is found. The Ridge position corresponds to the low-density structure, almost detached from the main body of the cloud and into which the shock is plunging (SiO at $\sim$39-40 km\,s$^{-1}$). Finally, the Ambient position is representative of dense unperturbed material within the IRDC. 

\begin{table}[!htpb]
    \centering
    \begin{tabular}{llllllllllllllllll}
    \hline
     \hline
         Position &RA(J2000) & Dec(J2000)\\
         & (hh:mm:ss) & (dd:mm:ss) \\
         \hline
        G3      &18:56:48.2 &01:23:40\\
        Shock   &18:56:42.3 &01:23:40\\
        Ridge   &18:56:43.4 &01:23:28\\
        Clump   &18:56:45.2 &01:23:52\\
        Ambient &18:56:46.0 &01:22:30\\
    \hline
    \hline
    \end{tabular}
    \caption{Equatorial Coordinates of the five positions analysed in this work.}
    \label{tab:tabpos}
\end{table}

Toward the five positions, we used the 30m single dish antenna at Instituto de Radioastronomia Millimetrica (IRAM-30m, Pico Veleta, Spain) to obtain single pointing N$_2$H$^+$(1-0) and N$_2$D$^+$(1-0) spectra in position switching mode (off position RA(J2000)=18$^h$57$^m$01$^s$, Dec(J2000)=1$^d$22$^m$25$^s$). The angular resolution of the 30m antenna is 34$^{\prime\prime}$ and 27$^{\prime\prime}$ at the N$_2$D$^+$ and N$_2$H$^+$ frequencies, respectively. These correspond to a linear spatial resolution of 0.4-0.5 parsec at the distance of G34.77 i.e., 2.9 kpc \citep{rathborne2006}. The FTS spectrometer was used with a frequency resolution of 200 kHz, corresponding to a velocity resolution between 0.7 and 0.8 km\,s$^{-1}$. Intensities were measured in units of antenna temperature, T$^{*}_{A}$, and converted into main-beam brightness temperature, $T_{\rm mb}$, using beam and forward efficiencies of 0.81 and 0.95, respectively. The final spectra were produced using the {\sc CLASS} software within the {\sc GILDAS} package\footnote{See http://www.iram.fr/IRAMFR/GILDAS.} and have final velocity resolution of 0.8~km~s$^{-1}$. The achieved rms per channel, $\sigma_{\rm rms}$, is 7~mK and 5 mK for the N$_2$H$^+$ and N$_2$D$^+$ spectra, respectively.

\section{Method}\label{method}

\begin{figure*}[!htpb]
    \centering
    \includegraphics[width=\textwidth,trim=3cm 0cm 3cm 0cm, clip=True]{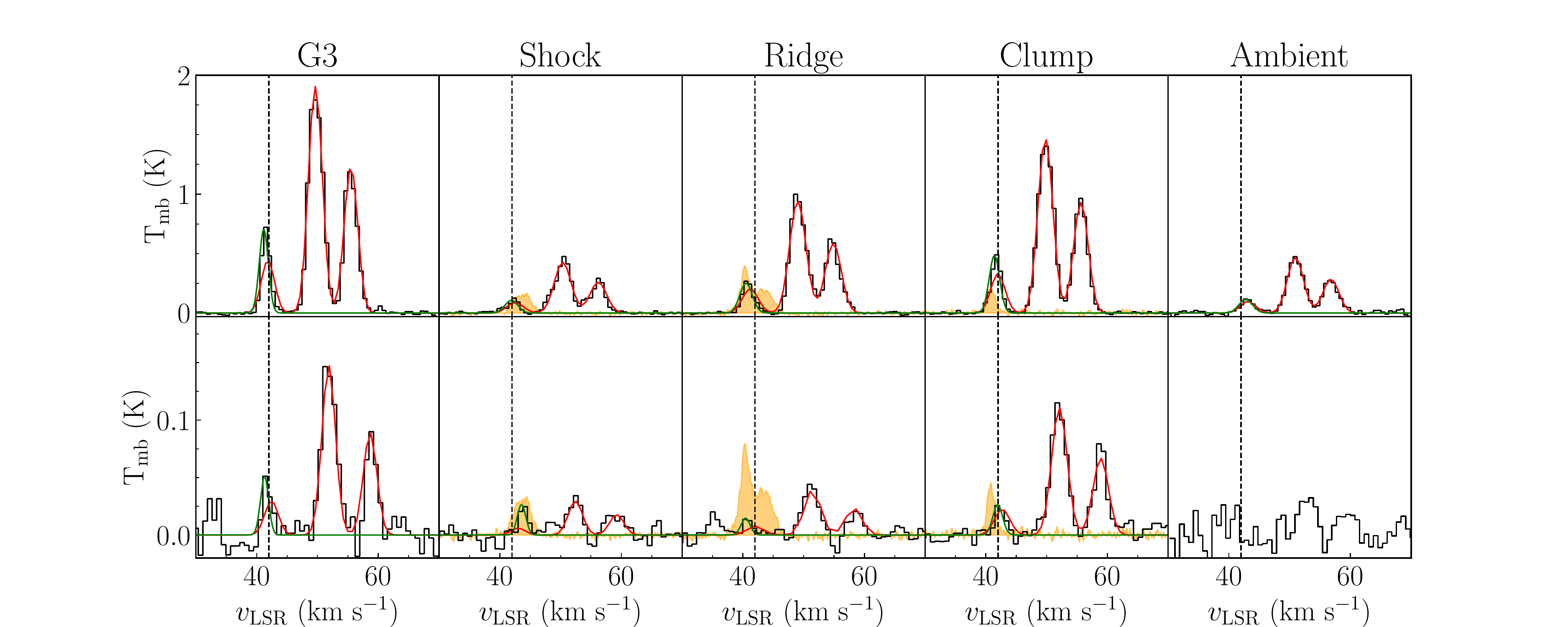}
    \caption{N$_2$H$^+$ (black curves, top panels) and N$_2$D$^+$ (black curves, bottom panels) obtained toward the five positions. The IRDC G34.77 central velocity is indicated as vertical dotted line in all panels. The red curves show the best LTE fitting models obtained by MADCUBA. The green curves show the gaussian fitting obtained for the isolated component. For the positions Shock, Ridge and Clump, we report the SiO(2-1) spectra (orange filled) extracted from the ALMA images \citep{cosentino2019} toward an angular region consistent with that of our current observations (34"). In the bottom panels the SiO spectra have been multiplied by 1/5 to allow a more straightforward comparison with the N$_2$D$^+$ spectra.}
    \label{fig:fig2}
\end{figure*}

The N$_2$H$^+$ (top panels) and N$_2$D$^+$ (bottom panels) spectra obtained toward the five positions are shown in Figure~\ref{fig:fig2}. The orange shadows in Figure~\ref{fig:fig2} correspond to the SiO(2-1) spectra obtained with ALMA in \cite{cosentino2019}. We detect significant N$_2$H$^+$ and N$_2$D$^+$ emission toward all positions except at the Ambient, where no N$_2$D$^+$ is observed above a threshold of 3$\times A_{\rm rms}$, where $A_{\rm rms}= \sigma_{\rm rms}\times dv \times\sqrt{N_{\rm channel}}$, with $dv$ the velocity resolution of the spectra and $N_{\rm channel}\simeq 4$ was estimated as the number of channels within a linewidth of 3 km s$^{-1}$, consistent with that estimated for both the N$_2$H$^+$ and N$_2$D$^+$ emission (see MADCUBA values in Table~\ref{tab:tabD1}). We note the presence of additional significant line emission at $\sim$35 km s$^{-1}$ within the N$_2$D$^+$ spectrum toward the Ridge. This component shows linewidth of $\sim$2.2 km s$^{-1}$ ($\sim$3 channels), peak intensity of $\sim$0.018 K and integrated intensity of $\sim$0.04 K km s$^{-1}$ i.e., $\sim$5$\times A_{\rm rms}$. We suggest that this line is either due to emission present along the line of sight or to an unknown species.\\
 
From the shown spectra, we estimate \deufrac as:
\begin{equation}
    D_{\rm frac}^{\rm N_2H^+} = \frac{N({\rm N_2D^+})}{N({\rm N_2H^+})},
    \label{eq1}
\end{equation}
\noindent
where $N$(N$_2$D$^+$) and $N$(N$_2$H$^+$) are the species total column densities. To estimate these quantities, we use the software MADCUBA\footnote{MADCUBA is a software developed in the Madrid Center of Astrobiology (INTA-CSIC). https://cab.inta-csic.es/madcuba/} \citep{martin2019} to fit the hyper-fine structure of the two species. By comparing the line emission with Local Thermodynamic Equilibrium (LTE) models, MADCUBA provides estimates of the species excitation temperature, $T_{\rm ex}$, column density, $N$, centroid velocity, $v_{\rm LSR}$, and linewidth, $\Delta v$. Initially, we fit the N$_2$D$^+$ and N$_2$H$^+$ spectra with the only assumption that the emission fills the beam (filling factor = 1). However, the software is unable to produce reasonable models, as indicated by the high uncertainties in the returned parameters (>200\%). This is likely due to the presence of a complex kinematic structure that is not well resolved at the velocity resolution of our observations (0.8 km s$^{-1}$). We thus conclude that it is not possible to estimate the excitation temperatures of the two species from a single rotational transition, and assume all lines to have the same $T_{\rm ex}$= 9 K, as measured from multiple CH$_3$OH transitions detected toward the shock peak in G34.77 \citep{cosentino2018}. How this assumption affects our results is discussed in  Appendix~\ref{appendixB}.

The best LTE fitting models are shown in Figure~\ref{fig:fig2} (red curves) for the five regions. In Table~\ref{tab:tab1}, $N$(N$_2$D$^+$) and $N$(N$_2$H$^+$) are reported together with the corresponding \deufrac values and the {\it Herschel}-derived mass surface densities, $\Sigma$, toward the five positions \cite{lim2016}. The Herschel-derived image has an angular resolution of 18$^{\prime\prime}$, but the $\Sigma$ values listed in Table~\ref{tab:tab1} have been extracted from a region with a 34$^{\prime\prime}$ aperture, consistent with that of the N$_2$D$^+$ observations.\\

We also estimate the N$_2$H$^+$ and N$_2$D$^+$ column densities from Equation~\ref{eq2}, i.e., Equation A4 from \cite{caselli2002}, under the assumption of optically thin emission: 
\begin{equation}
    N = \frac{8 \pi \nu^2}{A_{\rm ul} g_{u} c^3} \frac{k}{h} \frac{Q_{\rm rot}}{1-{\rm exp}(-h \nu /k T_{\rm ex})} \times \frac{{\rm exp}({E_{l}/(kT_{\rm ex})})}{J_{\nu}(T_{\rm ex})-J_{\nu}(T_{\rm bg})} \times W_{\rm tot}
    \label{eq2}
\end{equation}
\noindent
with the partition function, $Q_{\rm rot}$, defined as:
\begin{equation}
    Q_{\rm rot} = \displaystyle \sum_{J=1} ^{\infty} (2J+1) {\rm exp} \left(\frac{-E_J}{kT_{\rm ex}}\right)
    \label{eq3}
\end{equation}
\noindent
In Eq.~\ref{eq2} and \ref{eq3}, $\nu$ is the line frequency, $A_{\rm ul}$ is the Einstein coefficient for spontaneous emission and $g_{\rm up}$ is the statistical weight of the upper state. $J_{\nu}(T_{\rm ex})$ and $J_{\nu}(T_{\rm bg})$ are the equivalent Rayleigh-Jeans excitation and background temperatures ($T_{\rm bg}$=2.73 K) i.e., $J_{\nu}(T)$=$(h \nu / k) [exp(h \nu /kT)-1]^{-1}$. $J$ is the rotational state quantum number and $E_J= J(J+1)hB$ the corresponding state energy, calculated from the molecular rotational constant $B$. $W_{\rm tot}$ is the integrated intensity of the \textit{full} rotational transition $J=1-0$. Finally, $T_{\rm ex}$ is the line excitation temperature. The spectroscopic quantities for the two species have been obtained from the CDMS catalogue\footnote{See https://cdms.astro.uni-koeln.de/classic/.} and are reported in Table~\ref{tab:tabSpec}.

\begin{table}[!htpb]
    \centering
    \begin{tabular}{llllll}
    \hline
    \hline
         Species & Frequency & $E_{\rm up}$ & $g_{\rm up}$ & $A_{\rm ul}$ & $B$\\
                  & (GHz) & (K) & & (10$^{-5}$s$^{-1}$) & (MHz)\\
    \hline
         N$_2$D$^+$     &77.109    &3.70  &3 &2.06 & 38554.719\\
         N$_2$H$^+$     &93.173    &4.47  &3 &3.63 & 46586.867\\
    \hline
    \end{tabular}
    \caption{Spectroscopic information of the targeted species as obtained from the CDMS catalogue. }
    \label{tab:tabSpec}
\end{table}

We perform Gaussian fittings of the N$_2$H$^+$ and N$_2$D$^+$ isolated components and calculated the integrated intensity, $W_{\rm isolated}$, as the area underneath the Gaussian. Hence, we estimate  $W_{\rm tot}$ by scaling $W_{\rm isolated}$ for its relative intensity $R_i$=1/9. The best Gaussian fitting are reported in Figure~\ref{fig:fig2} (green curves), while in Table~\ref{tab:tab1}, we report the N$_2$H$^+$ and N$_2$D$^+$ column densities obtained from Eq~\ref{eq2} and assuming $T_{\rm ex}=9\:$K \cite{cosentino2018} and the associated \deufrac values. The best fitting parameters obtained with both the MADCUBA and Gaussian fitting analysis are reported in Table~\ref{tab:tabD1}.

\begin{table*}[]
    \centering
    \begin{tabular}{cc|ccc|ccc}
    \hline
     \hline
     & & \multicolumn{3}{c|}{MADCUBA} &\multicolumn{3}{c}{Gaussian Fitting}\\
         Position &$\Sigma$ & $N$(N$_2$H$^+$) & $N$(N$_2$D$^+$) & \deufrac & $N$(N$_2$H$^+$) & $N$(N$_2$D$^+$) &\deufrac\\
                  & (g cm$^{-2}$) &(10$^{13}$ cm$^{-2}$) & (10$^{12}$ cm$^{-2}$) & &(10$^{13}$ cm$^{-2}$) & (10$^{12}$ cm$^{-2}$) \\
         \hline
        G3      &0.09 &1.47$\pm$0.2 &1.33$\pm$0.2 &0.090$\pm$0.02 &1.55$\pm$0.16 &1.4$\pm$0.2  &0.09$\pm$0.01\\
        Shock   &0.04 &0.36$\pm$0.09 &0.26$\pm$0.1 &0.07$\pm$0.03 &0.34$\pm$0.04 &0.7$\pm$0.2 &0.22$\pm$0.06\\
        Ridge   &0.05 &0.80$\pm$0.2 &0.42$\pm$0.1 &0.053$\pm$0.02 &0.77$\pm$0.09 &0.4$\pm$0.1  &0.05$\pm$0.01\\
        Clump   &0.07 &1.17$\pm$0.1 &1.14$\pm$0.1 &0.097$\pm$0.02 &1.24$\pm$0.13 &0.9$\pm$0.2  &0.07$\pm$0.02\\
        Ambient &0.06 &0.40$\pm$0.09 &$\leq$0.1 &$\leq$0.03 &0.38$\pm$0.01 &$\leq$0.1 &$\leq$0.03\\
    \hline
    \hline
    \end{tabular}
    \caption{Mass Surface Density ($\Sigma$), N$_2$H$^+$ and N$_2$D$^+$ Column densities and Deuterium fraction, \deufrac, of the five positions analysed in this work using both the MADCUBA software and Gaussian fittings of the isolated components.}
    \label{tab:tab1}
\end{table*}

From Table~\ref{tab:tab1}, the \deufrac values obtained from the two methods (MADCUBA and Gaussian fitting of isolated components) are in agreement. The only significant discrepancy is found toward the Shock, where the \deufrac value estimated form the Gaussian fitting method ($\sim$0.2) is a factor of 3 larger than that obtained from the MADCUBA analysis ($\sim$0.07). Toward this position, the ratio between the intensity of the main and isolated N$_2$D$^+$ components is smaller ($\sim$1.2) than that seen toward other positions ($\sim$4.5). As a consequence, the N$_2$D$^+$ isolated component toward the Shock is not well reproduced by MADCUBA. This could be due to the isolated component being affected by a random noise spike that coincides with the line position, making it appear enhanced. Alternatively, the N$_2$D$^+$ emission toward the Ridge may need a non-LTE approach, for which the collisional coefficients of N$_2$D$^+$ are not available and multiple rotational transitions are needed. Finally, we exclude that the  N$_2$D$^+$ may be blended with another molecular species since no other molecular transitions are found at these frequencies.\\

Finally, besides the statistical uncertainty reported by MADCUBA for the N$_2$H$^+$ and N$_2$D$^+$ column densities, we also consider the systematic uncertainty due to the assumption $T_{\rm ex}$= 9 K. A $\pm$30\% variation of $T_{\rm ex}$, i.e., in the range 6-12 K, would result in a 5-10\% variation of the N$_2$H$^+$ column density and 3-10\% in the N$_2$D$^+$ column densities. Hence, we assume a systematic uncertainty of 10\% and add this in quadrature to the statistical errors. The obtained total uncertainties on the column density values are then propagated, with standard Gaussian rules, to estimate the uncertainties on the \deufrac. The final values are reported in Table~\ref{tab:tab1}.

\section{Results and Discussion}\label{resDisc}

\begin{figure}
    \centering
    \includegraphics[width=0.5\textwidth]{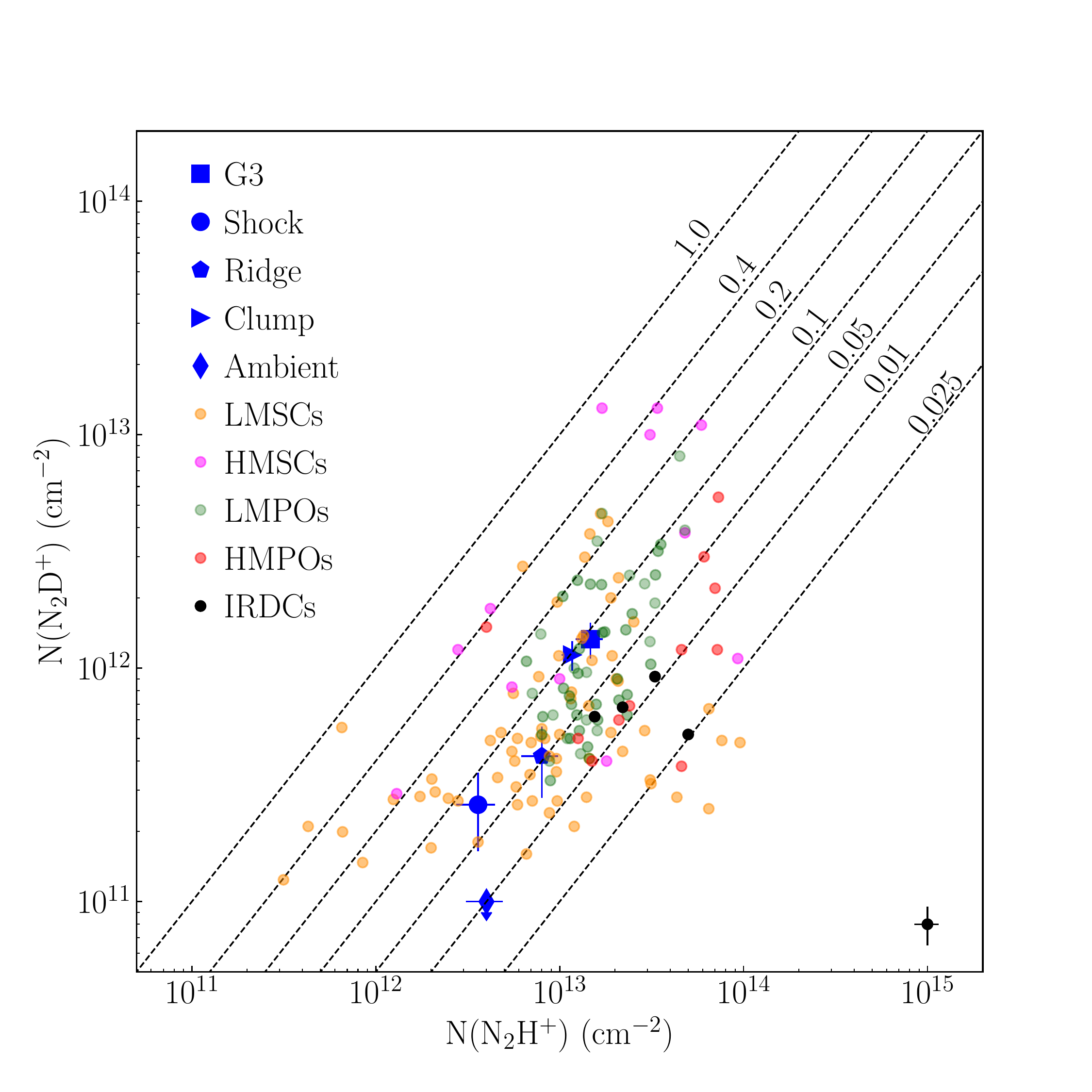}
    \caption{Column density values, from the MADCUBA analysis, of N$_2$D$^+$ as a function of N$_2$H$^+$ for all the positions analysed in this work, as well as literature values for Low-mass starless cores \cite[LMSCs;][]{crapsi2005,fontani2006,friesen2013,cheng2021}, high-mass starless cores \cite[HMSCs;][]{fontani2011,kong2016}, high-mass proto-stellar objects \cite[HMPOs;][]{fontani2011}, low-mass proto-stellar objects  \cite[LMPOs;][]{emprechtinger2009,friesen2013} and IRDCs \citep{miettinen2011,gerner2015,barnes2016}. 
    The average uncertainties associated to the data point from literature is reported in the bottom right corner.} Finally, dotted lines correspond to fixed values of D/H ratios.
    \label{fig:fig5}
\end{figure}

Toward all positions, we obtain D/H values several orders of magnitude larger than the cosmic D/H abundance. We find \deufrac in the range $\sim$0.05-0.1 toward G3, Shock, Ridge and Clump. For the Ambient gas, we estimate an upper limit of 0.03. These results are shown in Figure~\ref{fig:fig5}, where we compare the obtained \deufrac estimates with those typically observed in low-mass starless cores \cite[LMSCs;][]{crapsi2005,fontani2006,friesen2013,cheng2021}, high-mass starless cores \cite[HMSCs;][]{fontani2011,kong2016}, high-mass protostellar objects \cite[HMPOs;][]{fontani2011}, low-mass protostellar objects \cite[LMPOs;][]{emprechtinger2009,friesen2013} and large-scale regions of IRDCs \citep{miettinen2011,gerner2015,barnes2016}.

\begin{figure}[!htpb]
    \centering
    \includegraphics[width=0.5\textwidth, trim=0cm 0cm 0cm 0cm, clip=True]{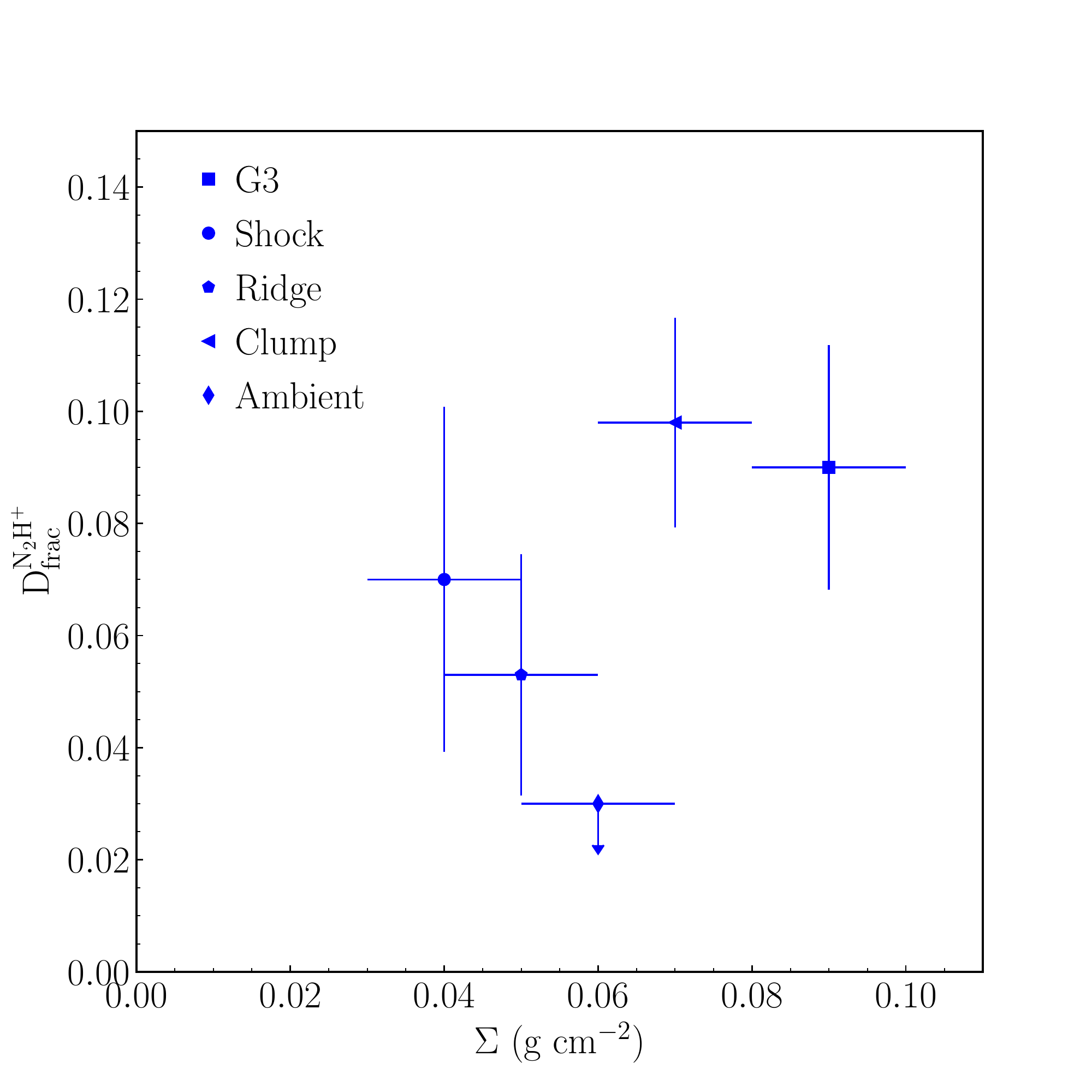}
    \caption{\deufrac obtained from the MADCUBA analysis, as function of $\Sigma$. Different symbols correspond to different positions. The uncertainty on the $\Sigma$ values is 0.01 g cm$^{-2}$, as reported by \cite{lim2016}}.
    \label{fig:fig3}
\end{figure}

In Figure~\ref{fig:fig3}, we show \deufrac as a function of mass surface density, $\Sigma$. From Figure~\ref{fig:fig3} and Table~\ref{tab:tab1}, the Shock and Ridge show lower or similar $\Sigma$ than the Ambient region. Hence, \deufrac would be expected to be lower or comparable to that measured toward the Ambient. However, an opposite trend is seen in Figure~\ref{fig:fig3}. Toward the Shock, the $N$(N$_2$H$^+$) is similar to that estimated for the Ambient gas, but the $N$(N$_2$D$^+$) is enhanced by at least a factor of 2. Toward the Ridge, the N$_2$H$^+$ column density is significantly larger than that reported toward the Ambient gas within the uncertainties, indicating the presence of more dense material toward this region. At the same time, the N$_2$D$^+$ column density toward this position is more than a factor of $\sim$3 larger than that measured for the Ambient gas. As a consequence, \deufrac toward the two positions is significantly enhanced, within the uncertainty, with respect to the unperturbed gas within the cloud. These differences indicate that, toward the Shock and the Ridge, additional processes may be boosting the production of D-bearing species with respect to the unperturbed Ambient gas.

\begin{table*}[]
    \centering
    \begin{tabular}{c|cc|cc|ccc|ccc}
    \hline
     \hline
     &\multicolumn{4}{c|}{MADCUBA} &\multicolumn{6}{c}{Gaussian Fitting}\\
     \hline
     &\multicolumn{2}{c|}{N$_2$H$^+$} &\multicolumn{2}{c|}{N$_2$D$^+$} &\multicolumn{3}{c|}{N$_2$H$^+$} &\multicolumn{3}{c}{N$_2$D$^+$}\\
         Position & $v_0$ & $\Delta v$ & $v_0$ & $\Delta v$ & $v_0$ & $\Delta v$  & $T_{\rm peak}$ & $v_0$ & $\Delta v$  & $T_{\rm peak}$\\
        & (km s$^{-1}$) & (km s$^{-1}$) &(km s$^{-1}$) & (km s$^{-1}$) &(km s$^{-1}$) &(km s$^{-1}$) & (K) &(km s$^{-1}$) &(km s$^{-1}$) & (K) \\
         \hline
        G3     &41.50$\pm$0.04  &2.5$\pm$0.1 &41.8$\pm$0.1 &2.6$\pm$0.2 &41.2 &1.7 &0.70 &41.3 &1.6 &0.05 \\
        Shock  &42.21$\pm$ 0.05 &3.2$\pm$0.1 &42.3$\pm$0.2 &2.7$\pm$0.5 &41.9 &2.6 &0.11 &43.6 &1.6 &0.03\\
        Ridge  &40.89$\pm$ 0.04 &3.1$\pm$0.1 &41.4$\pm$0.2 &3.1$\pm$0.4 &40.5 &2.4 &0.25 &40.4 &1.6 &0.02\\
        Clump  &41.67$\pm$ 0.03 &2.8$\pm$0.1 &42.08$\pm$0.04 &3.0$\pm$0.1  &41.4 &2.1 &0.48 &41.9 &1.9 &0.03\\
        Ambient &42.79$\pm$ 0.03 &3.1$\pm$0.1 &$\cdots$ &$\cdots$  &42.8 &2.6 &0.11 &$\cdots$  &$\cdots$  &$\cdots$\\
    \hline
    \hline
    \end{tabular}
    \caption{Best fitting parameters of the N$_2$H$^+$ and N$_2$D$^+$ emission lines obtained using MADCUBA and the Gaussian fitting of the isolated components.}
    \label{tab:tabD1}
\end{table*}

\begin{figure*}[]
    \centering
    \includegraphics[width=\textwidth, trim=2cm 18.5cm 3cm 4cm, clip=True]{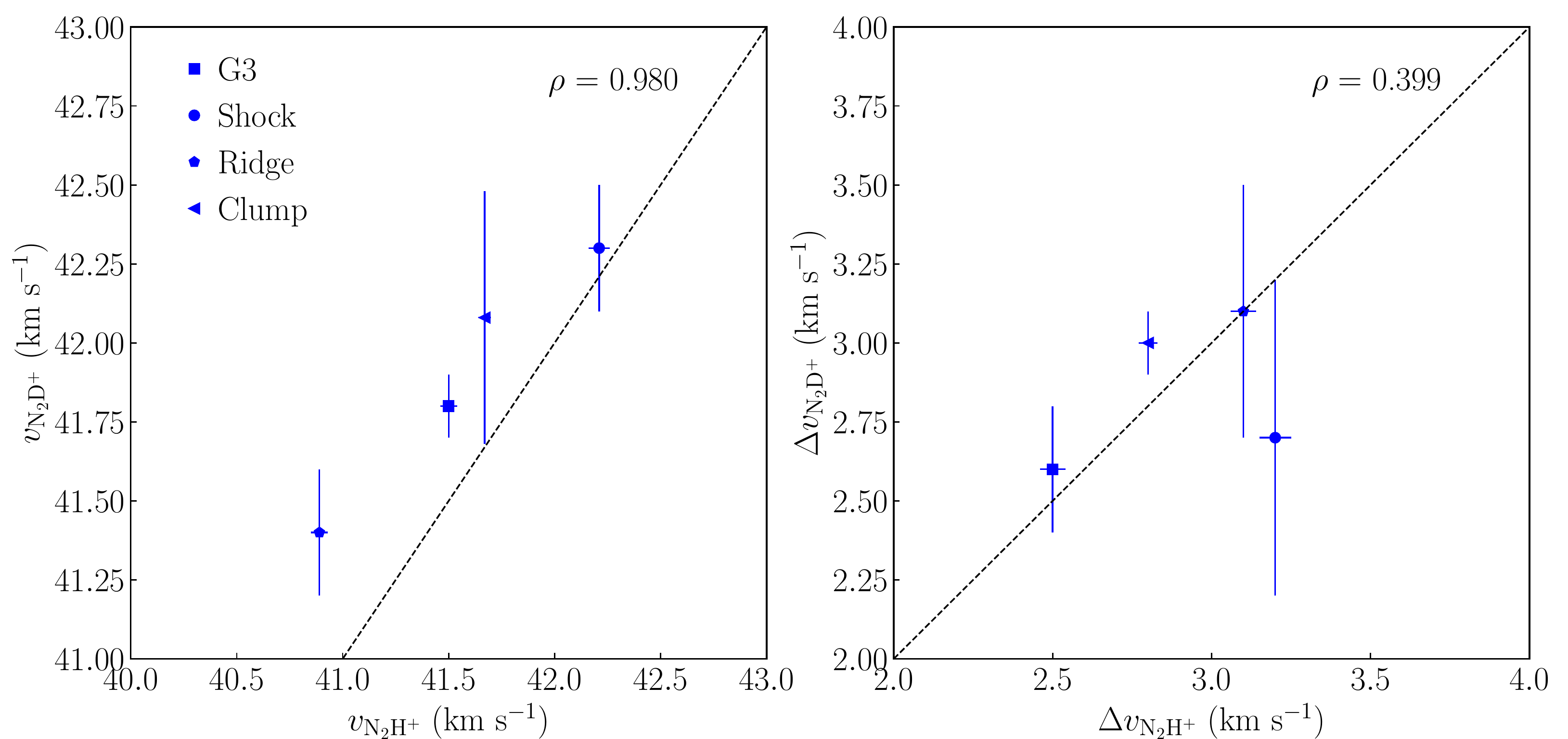}
    \caption{N$_2$D$^+$ centroid velocities (left) and linewidths (right) as a function of the corresponding N$_2$H$^+$ quantities. The Pearson's correlation coefficients are indicated within each panel.}
    \label{fig:figD1}
\end{figure*}

This enhanced \deufrac may be produced in the fast cooling post-shocked gas, where the shock has compressed the gas to densities $n$(H$_2$)$>$10$^5$ cm$^{-3}$. This high density is expected to shorten the CO depletion timescale and boost the production of N$_2$D$^+$ \citep{lis2002,lis2016}. For the post-shocked density $n$(H$_2$)$\geq$10$^5$ cm$^{-3}$ measured in \cite{cosentino2019}, the CO depletion timescale is expected to be $\leq$10$^4$ yr \citep{caselli1999}, consistent with the shock dynamical age \citep{cosentino2019}. On the other hand, as reported by \cite{codella2013} toward the molecular outflow shock L1157-B1, N$_2$H$^+$ is expected to be a fossil record of the pre-shocked gas that has been compressed. These two effects together result in an enhanced \deufrac toward the Shock and the Ridge. In accordance with this scenario, the CO depletion factor toward the two positions is in the range $\sim$4-6. CO depletion maps obtained toward G34.77 will be presented in a forthcoming paper (Petrova et al. in prep.) and have been obtained using $^{13}$CO(1-0) and C$^{18}$O(1-0) emission maps from the Green Bank Telescope. The $^{13}$CO(1-0) and C$^{18}$O(1-0) maps have been converted into column densities assuming excitation temperature of 7.5 K and by using the mass surface density maps derived from Herschel data \citep{lim2016}.\\ Finally, both the N$_2$H$^+$ and N$_2$D$^+$ centroid velocities toward the Ridge (Table~\ref{tab:tabD1}) are consistent with the velocity of the post-shocked gas \citep[$\sim40$ km s$^{-1}$; ][]{cosentino2019}. Toward the Shock, both the N$_2$H$^+$ and N$_2$D$^+$ emission show higher velocities, $\sim$42 km s$^{-1}$, but still consistent with the lowest velocities (and thus associated with the most downstream post-shocked gas), reached by the SiO toward that region (Figure~\ref{fig:fig2}). We note that the N$_2$H$^+$ column density toward the Ridge is a factor 1.5 larger than that measured toward the Shock. This may just reflect the decreasing amount of dense gas at the cloud outskirts (from the Ridge to the Shock) and/or a larger compression of the gas toward the Ridge. Indeed, here the shock velocities probed by SiO reach even lower values \citep[down to 39 km s$^{-1}$;][]{cosentino2019}.

The high \deufrac across the shock front may also be explained by the presence of a population of starless cores. In this parallel scenario, we speculate that the denser cooled-down post-shocked material is not diffuse, but it is organised in low-mass cores, whose formation may have been triggered by the shock passage that compresses gas to high-densities \cite{cosentino2019}.

Furthermore, as shown in Figure~\ref{fig:fig3}, the \deufrac estimates toward the Shock (blue circle) and the Ridge (blue plus) are consistent with those previously measured toward LMSCs \citep[orange circles][]{crapsi2005,fontani2006,friesen2013} and LMPOs \citep[green circles]{emprechtinger2009,friesen2013}. However, since no evidence of deeply embedded protostars is detected toward these regions \citep{cosentino2019,barnes2021}, we exclude that this putative low-mass population could have already reached the protostellar phase. 

We note that the timescale required for a core to reach \deufrac $\sim 0.1$ may be as short as 10$^4$ yr \citep{kong2015} under the physical conditions of density and cosmic ionisation rate similar to those estimated toward the shock in G34.77 \citep[n(H$_2$)$\sim$10$^5$ cm$^{-3}$ and $\zeta\sim$10$^{-15}$ s$^{-1}$;][]{cosentino2019} and if the H$_2$ ortho-to-para (OPR) ratio is $\leq$0.1. However, at the relatively low angular resolution of our observations it is not possible to distinguish between the above two scenarios of cooled-down diffuse post-shocked gas or population of low-mass pre-stellar cores. 

Toward the Clump, we measure \deufrac$\sim$0.09, consistent with that measured toward G3 and previously reported toward LMSC, HMSCs and LMPOs (Figure~\ref{fig:fig5}). The Clump also has a relatively high mass surface density of 0.07 g cm$^{-1}$ ($\sim60\:M_{\odot}$ within the 34$^{\prime\prime}$ beam), consistent with that of G3. Toward this position, \cite{barnes2021} report a N$_2$H$^+$ emission peak. Since the source shows no signatures of point-like emission at 4.5, 8 and 24 $\mu$m \citep{chambers2009} or 1 mm \cite{rathborne2006} and 3 mm continuum emission \citep{cosentino2019,barnes2021}, we exclude that the Clump may be hosting deeply embedded protostars. Due to its mass, density, lack of IR and mm continuum signatures, and high \deufrac value, we speculate that the Clump may still be evolving into the state of a centrally condensed pre-stellar core. However, it remains to be established if the gas is gravitationally bound and/or if it will fragment. We note that the Clump is located in the immediate vicinity of the shock front, where the shocked gas seen in SiO (Figure~\ref{fig:fig1}) has already decelerated down to 40 km s$^{-1}$. However, the N$_2$H$^+$ and N$_2$D$^+$ centroid velocities are not consistent with that of the post-shocked material. They indeed appear more in agreement with those measured toward the inner cloud, i.e., G3. Hence, we speculate that the Clump may have existed before the interaction with the SNR and has only been marginally affected by the shock.

\subsection{The N$_2$H$^+$ and N$_2$D$^+$ Line Profiles}

In Table~\ref{tab:tabD1}, we report the N$_2$H$^+$ and N$_2$D$^+$ centroid velocities, linewidths (Full Width Half Maxima) and line peaks obtained from the MADCUBA analysis and the Gaussian Fitting method, respectively. The two sets of parameters agree within the velocity resolution of our observations (0.8 km s$^{-1}$). In Figure~\ref{fig:figD1}, we show the correlations plots between the N$_2$H$^+$ and N$_2$D$^+$ centroid velocities (left) and linewidth (right), obtained from the MADCUBA analysis. We see that the centroid velocities of the two species are in agreement within the 0.8 km s$^{-1}$ velocity resolution. This is also the case for the linewidths of the species. The only exception may be represented by the Shock, where the N$_2$H$^+$ emission is slightly broader (3.2 km s$^{-1}$) than the corresponding N$_2$D$^+$ emission (2.7 km s$^{-1}$). In the proposed scenarios, this may be due to the fact that the N$_2$H$^+$ also probes gas in the act of being compressed, while the N$_2$D$^+$ does not. This putative complex kinematics are marginally seen in the velocity wings in the N$_2$H$^+$ spectrum, but cannot be fully resolved at the current velocity resolution. The presence of this additional N$_2$H$^+$ unresolved velocity component would result in the N$_2$H$^+$ column density being slightly overestimated. Hence, the \deufrac toward the Shock may be even larger than what is reported.

\section{Conclusions}\label{conclusions}

We have reported observations of the N$_2$H$^+$ and N$_2$D$^+$ $J=1-0$ emission toward five positions across the IRDC G34.77, namely a massive core, different positions across the shock front driven by the nearby SNR W44, a dense clump and unperturbed cloud material. We measured \deufrac across the cloud and compared the obtained results with those typically measured in star-forming regions at different evolutionary stages. We found that toward the G3, Shock, Ridge and Clump, \deufrac is significantly larger than that measured toward Ambient (\deufrac$\leq$0.03) and several orders of magnitude greater than the cosmic D/H ratio. We reported enhanced \deufrac toward both the Shock and the Ridge, where the gas is being compressed by the SNR-driven shock. Toward these regions, we suggest that the N$_2$H$^+$ emission may be a fossil record of the pre-shocked material that is now being compressed by the shock passage. We also suggest that the enhanced N$_2$D$^+$ column densities with respect to the Ambient region, toward these regions may be associated with the fast-cooling, much denser post-shocked material. Here, the high-densities previously reported enable a faster chemistry that quickly causes CO to be depleted and boosts the formation of D-bearing species in the gas phase. We also speculate that the shock passage may have triggered the formation of a low-mass starless core population in the dense post-shocked material. Here, the measured \deufrac may be produced toward these cores. High-angular resolution images are necessary to discern among the possible scenarios. Finally, we measured \deufrac$\sim$0.1 toward the Clump, similar to that observed toward starless cores, including G3. The source also has a mass surface density similar to that typically observed in clumps with the potential to harbour star formation. We speculate that this clump may represent a starless core in the making but we do not find evidence that this formation has been triggered by the shock passage.

\begin{acknowledgements}
G.C. acknowledges support from the Swedish Research Council (VR Grant; Project: 2021-05589). J.C.T. acknowledges support from ERC project 788829–MSTAR. I.J-.S acknowledges funding from grant No. PID2019-105552RB-C41 awarded by the Spanish Ministry of Science and Innovation/State Agency of Research MCIN/AEI/10.13039/501100011033. JDH gratefully acknowledges financial support from the Royal Society (University Research Fellowship; URF/R1/221620). P.G. acknowledges support from the Chalmers Cosmic Origins postdoctoral fellowship. R.F. acknowledges funding from the European Union’s Horizon 2020 research and innovation programme under the Marie Sklodowska-Curie grant agreement No 101032092. R.F. also acknowledges support from the grants Juan de la Cierva FJC2021-046802-I, PID2020-114461GB-I00 and CEX2021-001131-S funded by MCIN/AEI/ 10.13039/501100011033 and by ``European Union NextGenerationEU/PRTR''. S.V. acknowledges partial funding from the European Research Council (ERC) Advanced Grant MOPPEX 833460. 
\end{acknowledgements}

\bibliographystyle{aa} % style aa.bst
\bibliography{aa.bib} % your references Yourfile.bib

\begin{appendix}\label{appendixA} %First appendix
\section{The HFS GILDAS Fitting Method}\label{appendixB}

The HFS method in {\sc GILDAS} is a fitting procedure that reproduces the N$_2$H$^+$ and N$_2$D$^+$ hyper-fine structures in the regime of optically thin and thick lines. In the case of optically thick lines, the excitation temperature can be estimated, together with the total line optical depth and the centroid velocity. We have applied the HFS method to the N$_2$H$^+$ and N$_2$D$^+$ spectra in Figure~\ref{fig:fig2}. For all the N$_2$D$^+$ spectra, the HFS method indicates that the lines are optically thin toward all positions. It is thus not possible to estimate $T_{\rm ex}$ for this species. For the N$_2$H$^+$ spectra, the HFS method indicates optically thin lines toward all position except the core G3. Furthermore, toward the Shock, Ridge and Clump, the method also returns a bad fit. This is likely due to the presence of a complex kinematic structures not resolved at the low velocity resolution of our observations. As a consequence, it is not possible to estimate $T_{\rm ex}$ of N$_2$H$^+$ toward these positions. The only spectrum for which the HFS is successful is the N$_2$H$^+$ emission toward the core G3, for which we report the best fitting parameters in Table ~\ref{tab:tabC1}. 

\begin{table}[!htpb]
    \centering
    \begin{tabular}{lllll}
    \hline
    \hline
         Position & $T_{A}\times\tau$ & $v_{\rm lsr}$ & $\Delta v$  & $\tau_{\rm main}$\\
         (GHz) & (K) & (km s$^{-1}$) &(km s$^{-1}$) &\\
    \hline
G3    &5.5$\pm$0.1   &41.600$\pm$0.005   &2.1$\pm$0.1     &1.48$\pm$0.1\\
    \hline
    \end{tabular}
    \caption{Best fitting parameters from the HFS method in GILDAS applied to the N$_2$H$^+$ spectra toward the core G3.}
    \label{tab:tabC1}
\end{table}

\noindent
From these parameters, we estimate the excitation temperature using the following equation
\citep[e.g.,][]{punanova2016}:
\begin{equation}
    T_{\rm ex} = \frac{h\nu}{k} \left[ ln \left( \frac{(h\nu/k)}{(T_{A} \times \tau)/\tau + J_{\nu}(T_{\rm bg})} +1 \right) \right]^{-1}
    \label{eq:eq4}
\end{equation}
\noindent
where $\nu$ is the frequency of the component used as reference in the HFS fitting; J$_{\nu}$($T_{\rm bg}$) is the equivalent Rayleigh-Jeans background temperature ($T_{\rm bg}$=2.73 K); $\tau$ is the optical depth of the main component. In the case of optically thick lines, $T_{A}\times \tau$ is the total optical depth times the difference between the Rayleigh-Jeans equivalent excitation and background temperatures, while for the case of optically thin lines it corresponds to the main beam temperature of the main component. From Equation~\ref{eq:eq4}, we estimate the excitation temperature to be $T_{\rm ex}$=7 K. This value is consistent with that reported in other low-mass \citep{friesen2013} and high-mass cores \citep{fontani2011,kong2016} and toward other IRDCs \citep{barnes2016}. Assuming $T_{\rm ex}$=7 K would lead to \deufrac values lower by less than 10\% with respect to those reported in Table~\ref{tab:tab1} for $T_{\rm ex}=9\:$K.

\end{appendix}
\end{document}